\documentclass{amsart}
\usepackage{amsmath}
\usepackage{amssymb}
\usepackage{amsthm}
\usepackage{natbib}

\theoremstyle{definition}
\newtheorem{definition}{Definition}[section]
\newtheorem{lemma}{Lemma}[section]
\newtheorem{theorem}{Theorem}[section]
\DeclareMathOperator{\arcsinh}{arcsinh}
\DeclareMathOperator{\disc}{Disc}
\DeclareMathOperator{\Real}{Re}
\DeclareMathOperator{\Imag}{Im}

\allowdisplaybreaks

\newcommand{\EVAL}{\texttt{SqFreeEVAL}}
\newcommand{\EVALTYPE}{\texttt{EVAL}-type}
\newcommand{\GEN}{\texttt{Bisection}}
\DeclareMathOperator{\HM}{HM}

\begin{document}

\title{\EVAL: An (almost) optimal real-root isolation algorithm}

\author{Michael A. Burr}
\address{Fordham University, 441 East Fordham Road, Bronx, NY 10458, USA}
\email{mburr1@fordham.edu}

\author{Felix Krahmer}
\address{Hausdorff Center for Mathematics, Universit\"at Bonn, Endenicher Allee 60, 53115 Bonn, Germany}
\email{felix.krahmer@hcm.uni-bonn.de}

\begin{abstract}
Let $f$ be a univariate polynomial with real coefficients, $f\in\mathbb{R}[X]$.  Subdivision algorithms based on algebraic techniques (e.g., Sturm or Descartes methods) are widely used for isolating the real roots of $f$ in a given interval. In this paper, we consider a simple subdivision algorithm whose primitives are purely numerical (e.g., function evaluation).  The complexity of this algorithm is adaptive because the algorithm makes decisions based on local data.  The complexity analysis of adaptive algorithms (and this algorithm in particular) is a new challenge for computer science. In this paper, we compute the size of the subdivision tree for the \EVAL\ algorithm.

The \EVAL\ algorithm is an evaluation-based numerical algorithm which is well-known in several communities.  The algorithm itself is simple, but prior attempts to compute its complexity have proven to be quite technical and have yielded sub-optimal results.  Our main result is a simple $O(d(L+\ln d))$ bound on the size of the subdivision tree for the \EVAL\ algorithm on the benchmark problem of isolating all real roots of an integer polynomial $f$ of degree $d$ and whose coefficients can be written with at most $L$ bits.

Our proof uses two amortization-based techniques: First, we use the algebraic amortization technique of the standard Mahler-Davenport root bounds to interpret the integral in terms of $d$ and $L$.  Second, we use a continuous amortization technique based on an integral to bound the size of the subdivision tree.  This paper is the first to use the novel analysis technique of continuous amortization to derive state of the art complexity bounds.

\end{abstract}

\maketitle

\section{Introduction}
In this paper, we show that the size of the subdivision tree for the simple, evaluation-based, numerical algorithm \EVAL\ has size $O(d(L+\ln d))$ for the benchmark problem of isolating all of the real roots of an integer polynomial of degree $d$ whose coefficients can be represented by at most $L$ bits.  Under the mild assumption that $L\geq\ln d$, this complexity simplifies to the optimal size of $O(dL)$.  The optimality and simplicity of the \EVAL\ algorithm imply that it may be a useful algorithm in practical settings.  The bound on the size of the subdivision tree is achieved via a straight-forward and elementary argument.  The two main techniques which are used in the computation are algebraic amortization, in the form of Mahler-Davenport bounds, and continuous amortization, in the form of an integral technique as presented in \citep{Burr-Krahmer-Yap:integral:09}.

\subsection{\EVALTYPE\ algorithms}
The \EVAL\ algorithm which we study in this paper is a specific example of what we call an \EVALTYPE\ algorithm.  These algorithms are so named because they are based on function evaluation: \EVALTYPE\ algorithms take, as input, functions which allow some subset of the following two predicates: First, these functions and their derivatives can be evaluated at a countable dense subset of their domain.  In this paper, the domain will be the real numbers and the countable dense subset will be the dyadic integers.  Second, these functions and their derivatives can be approximated on intervals in such a way that the approximation converges as the input intervals converge to a point.  In this paper, the approximation is derived from interval arithmetic on a Taylor sequence.  The simplest and most well-known example of an \EVALTYPE\ algorithm is Lorensen and Cline's marching cube algorithm \citep{marchingcube1987}.

\EVALTYPE\ algorithms are typically studied because of their simplicity and generality.  These algorithms are fairly general because their inputs can be extended to more general analytic functions.  In particular, many analytic functions have interval arithmetic available to them, and, therefore, it is possible to approximate these functions on intervals.  In addition, with the limited predicates available to \EVALTYPE\ algorithms, most of the techniques which are used in these algorithms are analytically based (as opposed to algebraically based).  These algorithms are simple because, in many cases, \EVALTYPE\ algorithms are based on simple recursive bisection algorithms.  Such algorithms iteratively subdivide an initial domain until each set in the resulting partition of the initial domain satisfies a (usually simple) terminal condition.  Bisection algorithms are common in computer graphics \citep{SurveySubdivision} as well as in computational science and engineering applications \citep{domaindecomposition}.  Bisection algorithms are of particular interest because they are adaptive; they perform more bisections near difficult features and fewer bisections elsewhere.  However, this adaptivity makes the complexity analysis of such algorithms more difficult because the subdivision tree may have a few deep paths while the remainder of the tree remains modest in size.

\EVALTYPE\ algorithms have been studied in the univariate case in \citep{Henrici:search:70,Yakoubsohn:bisection:05,sagraloff-yap:ceval:09,burr-sharma-yap:eval:09,Burr-Krahmer-Yap:integral:09}, in the bivariate and trivariate cases in \citep{marchingcube1987,snyder:interval:92,plantinga-vegter:isotopic:04,plantinga:thesis:06,linyap2009,burr+3:subdiv2:10}, and in the multivariate case in \citep{galehouse:thesis}.  All of these algorithms are devoted to approximating algebraic (and in some cases analytic varieties) in the real or complex settings.  The algorithms in \citep{burr-sharma-yap:eval:09,Burr-Krahmer-Yap:integral:09} are designed to find all real roots of a polynomial or analytic function while the algorithms in \citep{Henrici:search:70,Yakoubsohn:bisection:05,sagraloff-yap:ceval:09} are designed to find the complex roots of a polynomial or analytic function (note that \citep{Henrici:search:70} is only designed to find a single root of a polynomial).  Each of these algorithms is very closely related to the \EVAL\ algorithm considered in this paper; the main differences are in the setting, in the type of subdivisions performed, and in various preprocessing steps.  We give a more detailed account of these algorithms in the next section.  The two-dimensional \EVALTYPE\ algorithm \citep{plantinga-vegter:isotopic:04,plantinga:thesis:06} was presented for approximating smooth and bounded varieties.  It was extended to singular and unbounded varieties in \citep{burr+3:subdiv2:10}; in addition, the tests performed by the algorithm were improved in \citep{linyap2009}.

\subsection{The \EVAL\ algorithm}
There are many bisection algorithms for finding roots, see Section \ref{Intro:RootsFinding} for references, but among such algorithms, the \EVAL\ algorithm is one of the simplest and most widely applicable, see \citep{burr-sharma-yap:eval:09}.  There are two distinct paths in the literature which arrive at algorithms similar to the \EVAL\ algorithm: one path proceeds through the consideration of magnitudes of derivatives and the other path proceeds via interval arithmetic.

We begin by discussing the history from the magnitudes of derivatives perspective.  In \citep{Henrici:search:70}, the author presents an algorithm for finding a single complex root of a polynomial.  The test $T_3$ from the paper is essentially used here.  In \citep{Yakoubsohn:bisection:05}, the test is developed into a bisection algorithm and to find all complex roots of entire functions, not just polynomials.  In the paper, however, the test from \citep{Henrici:search:70} is used only as a one-sided test; therefore, the algorithm can only exclude regions from containing roots and does not confirm that roots exist in the final regions.  There, the algorithm was termed a {\em bisection-exclusion} algorithm to reflect this drawback.  Finally, in \citep{sagraloff-yap:ceval:09}, the algorithm from \citep{Yakoubsohn:bisection:05} was adapted to polynomials in order to confirm that roots exist in the final regions; there, the authors studied both an algorithm for finding complex roots as well as one for finding real roots.  The \EVAL\ algorithm is a natural restriction of these complex root-finding algorithms to the real line.

On the other hand, from the interval arithmetic community, a bisection algorithm using interval methods was suggested in \citep{moore:bk,mitchell:robust-ray:90}.  In these papers, any interval function can be used; if the standard centered form for polynomials is used, see \citep{ratschek-rokne:range:bk}, then the exclusion conditions are identical (when $f$ and $f'$ are square free) to those for the \EVAL\ algorithm.

In this paper, we study the \EVAL\ algorithm on the standard benchmark problem of finding all of the real roots of a polynomial.  We show that, in this case, the subdivision tree has the favorable size of $O(d(L+\ln d))$ which simplifies to the optimal size of $O(dL)$ under the mild assumption that $L\geq\ln d$.  Since this algorithm uses only local data to find roots, it is an adaptive algorithm and may be more efficient than the standard exact algorithms in certain cases, see \citep{burr-sharma-yap:eval:09}.  In addition, the \EVAL\ algorithm can handle analytic varieties, see \citep{burr-sharma-yap:eval:09}, which extends its reach beyond that of more standard exact algorithms which require sophisticated algebraic primitives and are specialized to polynomials.  These advantages of the \EVAL\ algorithm imply that it may be more practical than other standard root isolation algorithms in practice.

\subsection{Previous complexity results}

The computational complexity of \EVALTYPE\ algorithms has proven to be quite a challenging problem because the algorithms are adaptive and the analytic primitives do not carry much information about the global structure (unlike algebraic information).  Here, we survey the complexity analyses of the precursors to the \EVAL\ algorithm.  In most situations, the complexity is computed in terms of the size of the subdivision tree of the specific \EVALTYPE\ algorithm (this is almost equivalent to counting the number of tests performed by the algorithm). There have been two main techniques to find the size of the subdivision tree: by finding the width of the subdivision tree at various subdivision levels or by finding the local depth of the subdivision tree.

In \citep{Henrici:search:70}, the author is searching for only a single root, and, therefore, retains a single disk containing a root at each stage of the algorithm.  Many tests are performed in the algorithm, however, because at each stage of the algorithm, tests are performed on a covering of the previously retained disk. The final stopping criterion for this algorithm is based on a precision $\epsilon>0$ which is chosen {\em a priori} by the user.  When the worst-case root separation bound for a polynomial is used, the complexity of the subdivision tree becomes $O(d^3(L+\ln d))$.

In \citep{Yakoubsohn:bisection:05}, the author is searching for all of the complex roots of an analytic function.  In the computation, a bound on the width of the tree is computed to bound the number of subdivisions performed.  Since this algorithm only excludes regions and lacks an inclusion test, it is possible that the final output regions do not contain roots or do not separate roots.  The final stopping criterion for this algorithm is based on a precision $\epsilon>0$ which is chosen {\em a priori} by the user. When the worst-case root separation bound for a polynomial is used, the complexity of the subdivision tree becomes either $O(d^4(L+\ln d))$ or $O(d^3(L+\ln^3 d))$ after $\lceil\ln d\rceil$ steps of the Graeffe iteration.

In \citep{Burr-Krahmer-Yap:integral:09}, we search for all of the real roots of a polynomial.  Here, the computation is based on the depth of the tree over each point of the initial interval.  In the paper, we introduced the idea of continuous amortization via an integral and showed how to use it to bound the size of the subdivision tree.  In particular, we proved a complexity bound of $O(d^3(L+\ln d))$ for the subdivision tree.

In \citep{sagraloff-yap:ceval:09}, the authors present algorithms to find all of the real or all of the complex roots of a polynomial.  In the computation, a bound on the width of the subdivision tree is used to compute the number of subdivisions performed.  The authors show that the complexity of the subdivision tree is $O(d(L+\ln d)(\ln L+\ln d))$ in the real case and $O((d\ln d)^2(L+\ln d))$ in the complex case.  In addition, the authors show that the bit complexity of both of these algorithms is $\widetilde{O}(d^4L)$ where the $\widetilde{O}$ means that logarithmic factors in $d$ and $L$ have been suppressed.

Each of the analyses in \citep{Yakoubsohn:bisection:05,Burr-Krahmer-Yap:integral:09,sagraloff-yap:ceval:09} are quite technical, complicated, and require several constants to be defined whose use becomes justified only after the completion of the complexity analysis.  In contrast, the computation in this paper is quite simple and provides the better bound of $O(d(L+\ln d))$.  It should be noted that although this is the best bound known, it does not directly replace the bounds presented in these papers because some are in the different setting of the complex plane and others use different preprocessing steps.  In the case where the polynomial and its derivative are both square free and we are searching for the real roots, all of these algorithms are identical and our bound on the subdivision tree is the best.

\subsection{Algebraic and continuous amortization}
In this paper, we use amortization in two forms: algebraic and continuous.  Algebraic amortization originated with Davenport \citep{davenport:85} where the individual root separation bounds are replaced by a product of root separations.  This bound was then studied in \citep{du-sharma-yap:sturm:07,eigenwillig-sharma-yap:descartes:06} where it was generalized to other root separation products including complex roots.  This technique has proven useful to compute the complexity of the subdivision tree for many other root isolation techniques, see Section \ref{Intro:RootsFinding}.  We introduced continuous amortization in \citep{Burr-Krahmer-Yap:integral:09} to bound the size of the subdivision tree of an \EVALTYPE\ algorithm.  In this paper, we show that continuous amortization can be used to significantly simplify complexity calculations.

In continuous amortization, we use a complexity charge $\phi$ whose domain is the input region, and, for each $x$ in the input region, $\phi(x)$ is a lower bound on the size of any leaf interval containing $x$.  Then $1/\phi(x)$ is related to the depth of the subdivision tree for an interval which contains $x$.  In \citep{Burr-Krahmer-Yap:integral:09}, we used continuous amortization to compute the size of a subdivision tree for an \EVALTYPE\ algorithm.  In this paper, we greatly simplify the computation and provide a complexity bound for the \EVAL\ algorithm.

We call the function $\phi$ mentioned in the previous paragraph a {\em stopping function} for the algorithm.  Similar functions also appeared in \citep{Henrici:search:70} where they were called {\em inner} and {\em outer convergence} functions.  In \citep{Yakoubsohn:bisection:05} such functions were also termed {\em exclusion functions}.  In both cases, these stopping functions were used to compute the complexity of the algorithm, but they were not used in a continuous amortization computation.

\subsection{Other root isolation algorithms}\label{Intro:RootsFinding}
There is an extensive amount of literature on the complexity of root isolation, see \citep{pan:history-progress:97,Pan1996} for surveys of the previous literature, which we will not attempt to cover here.  Most algorithms are compared by their performance on the benchmark problem of finding all real roots of a polynomial of degree $d$ and whose coefficients can be represented by at most $L$ bits.  For this problem, the bit-complexity of $O(d^3(L+\ln d))$ for complex roots
was first achieved by Sch\"onhage \citep{schonhage:fundamental}.  In many algorithms, the size of the subdivision tree is smaller than this bound because, for each node in the subdivision tree, additional calculations must be performed.  Davenport \citep{davenport:85} proved that the the subdivision tree for the Sturm method is $O(d(L+\ln d))$, see \citep{reischert:subresultant:97,lickteig-roy:sequences:01,du-sharma-yap:sturm:07}.  More recently, it has been shown in \citep{eigenwillig-sharma-yap:descartes:06} that the Descartes method also achieves this bound, see \citep{collins-akritas:76,eigenwillig-sharma-yap:descartes:06,krandick-mehlhorn:06,collins-johnson-krandick:cad:02}.  These methods are optimal under the weak assumption that $L\geq\ln d$.  In addition, related exact techniques using continued fractions were shown to have a tree size of $\widetilde{O}(dL)$ when an ideal root bound is used and $\widetilde{O}(d^2L)$ when a more practical bound is used \citep{sharma:2008}. In the algebraic computing community, the Descartes method appears to be one of the more practical algorithms, see \citep{collins-johnson-krandick:cad:02,johnson:root-isolation:98,rouillier-zimmermann:roots:04,mrr:bernstein:05,rouillier-zimmermann:roots:04}.  In this paper, we show that the subdivision tree for the \EVAL\ algorithm also achieves this bound; therefore, the \EVAL\ algorithm should also be considered on equal footing with the other more well-known root finding algorithms via the Sturm or Descartes methods.  The \EVAL\ algorithm may, in addition, be considered practical because its computations are numerical and hence easy to implement and its subdivision tree has a favorable size.

\subsection{Organization of this paper}
In Section \ref{Sec:EVAL}, we introduce the \EVAL\ algorithm and discuss the main condition we will use for an interval to be \EVAL\ terminal.  In Section \ref{Sec:Integral}, we review the use of stopping functions to bound the size of the subdivision from \citep{Burr-Krahmer-Yap:integral:09} and create a stopping function for the \EVAL\ algorithm. In Section \ref{Sec:Analysis}, we compute the size of the \EVAL\ algorithm's subdivision tree using continuous amortization via the stopping function technique and achieve the main result of this paper, the $O(d(L+\ln d))$ bound on the size of the subdivision tree for the \EVAL\ algorithm.  Finally, we conclude in Section \ref{Conc}.

The authors would like to thank the following people for many useful discussions: Benjamin Galehouse, Michael Sagraloff, and Chee Yap.

\section{The \EVAL\ algorithm}\label{Sec:EVAL}
Given an interval $I=[a,b]$ with integer endpoints and a polynomial $f$ with integer coefficients, i.e., $f\in\mathbb{Z}[X]$, the \EVAL\ algorithm returns a collection of intervals which cover and isolate the real roots of $f$ in $(a,b)$, i.e., every root appears in an output interval and each output interval contains exactly one root (ignoring multiplicities).  In the \EVAL\ algorithm, if the interval $[c,d]$ is output, then $(c,d)$ contains exactly one root of $f$ and if $[c,c]$ is output, then $c$ is a root of $f$.  The \EVAL\ algorithm maintains a (finite) partition $P$ of the interval $I$, i.e., a finite collection of intervals whose interiors are disjoint and whose union is $I$.  The \EVAL\ algorithm iteratively bisects the elements of $P$ until the intervals of the partition $P$ are each small enough to pass the \EVAL\ termination conditions (see Section \ref{Subsec:EVAL}).  Of interest to us is the size $\#P$ of the partition, i.e., the number of intervals in $P$.

We begin with some terminology: For an interval $J=[c,d]$ the {\em width} of $J$ is $w(J)=d-c$ and the {\em midpoint} of $J$ is $m(J)=(c+d)/2$.  Also, to {\em bisect} an element of the partition $P$ means to replace the interval $J=[c,d]\in P$ by the two subintervals $[c,m(J)]$ and $[m(J),d]$.  Note that this implies that $\#P$ is one more than the number of bisections done by the \EVAL\ algorithm, i.e., the size of the subdivision tree. All of the calculations done by the \EVAL\ algorithm will be performed on the dyadic integers $\mathbb{Z}[1/2]$ so that all of the standard operations are exact.  This prevents well-known implementation errors from arising in practice.
\subsection{Statement of the \EVAL\ algorithm}\label{Subsec:EVAL}
In the \EVAL\ algorithm, we first replace $f$ by its square free component, which we briefly call $g$.  Then, we replace $f'$ by its square free and relatively prime to $f$ component, i.e., we first take the square free component of $f'$ and then take the portion of this polynomial which is relatively prime to $f$.  We briefly call this $h$.  Note that $g|f$ and $h|f'$, and, moreover, the roots of $g$ are separated by roots of $h$ by Rolle's theorem.  In the case where $f$ is square free, the zeros of $h$ partition $f$ into monotonic regions; in the case where $f$ is not square free, the zeros of $h$ no longer have this property, but they still partition the roots of $f$ (and hence the roots of $g$).  Throughout the remainder of this paper, except for a brief note in Section \ref{subsec:EVALbounds}, we use these square free substitutions for $f$ and $f'$ without mention.  The bounds on the subdivision tree, however, will be in terms of the data for original the $f$ and not for any replacements.

The \EVAL\ algorithm creates a partition of $I$ and determines which intervals in the partition contain roots.  Initially, the partition of $I$ is $P=\{I\}$, the trivial partition.\\[.2cm]
\fbox{\parbox{4.7in}
{\begin{center}
Algorithm 2.1: The \EVAL\ algorithm
\end{center}
\it
Repeatedly subdivide each $J\in P$ until one of the following conditions holds:\\[.15cm]
\hspace*{20pt} \textnormal{($\text{C}_0$)}\qquad$\displaystyle |f(m(J))|>\sum_{i=1}^d\frac{|f^{(i)}(m(J))|}{i!}\left(\frac{w(J)}{2}\right)^i$ or \\
\hspace*{20pt} \textnormal{($\text{C}_1$)}\qquad$\displaystyle |f'(m(J))|>\sum_{i=1}^{d-1}\frac{|f^{(i+1)}(m(J))|}{i!}\left(\frac{w(J)}{2}\right)^i$\\[.15cm]
\hspace*{20pt}If, when subdividing, $f(m(J))=0$, then output $[m(J),m(J)]$.\\[.15cm]
For each interval $J=[c,d]\in P$ where \textnormal{$\text{C}_1$} holds and $f(c)\cdot f(d)<0$, output $J$
}}\\

The termination proof for the \EVAL\ algorithm is very similar to the corresponding statement in \citep{Burr-Krahmer-Yap:integral:09,sagraloff-yap:ceval:09}.  The correctness proof is slightly different from the corresponding proofs for other \EVALTYPE\ algorithms.  The correctness follows from the Taylor polynomial centered at $m(J)$: if one of the conditions holds, then it follows that $f$ (for condition $C_0$) or $f'$ (for condition $C_1$) is never zero in $J$ since the inequalities are equivalent to a reverse triangle inequality on the Taylor polynomial.  The first condition implies that $f$ has no zeros in $J$.  The second condition implies that $f$ has at most one zero in $J$ since roots of $f'$ separate zeros of $f$ (even though $f$ might not be monotonic due to the replacements above).

\subsection{\EVAL\ terminal intervals}\label{sec:evaltermint}
In this section, we provide a sufficient condition for the \EVAL\ algorithm to terminate without subdividing on a given interval, i.e., for the interval to be \EVAL\ terminal.

\begin{definition}
For any polynomial $g$ of degree $d$, define $\alpha_g=\{\alpha_1,\cdots,\alpha_d\}$ to be the multiset of the roots of $g$. In addition, define the function $\Sigma_g$ to be the sum of the reciprocals of the distances from its argument to the roots of $g$:
$$\Sigma_g(x)=\sum_{\alpha\in \alpha_g}\frac{1}{|x-\alpha|}.$$
Note that this function can be represented in a simple form using the harmonic mean $\HM$.  Then, one has
$$\frac{1}{\Sigma_g(x)}=\frac{\HM(|x-\alpha_g|)}{d}$$
where $|x-\alpha_g|$ is the set of distances from $x$ to the roots of $g$.
\end{definition}
$\Sigma_f$ and $\Sigma_{f'}$ will be our main objects of study.  We begin with the following lemma which connects $\Sigma_f(x)$ and $\Sigma_{f'}(x)$ with conditions $C_0$ and $C_1$, respectively:
\begin{lemma}\label{lem:Sigma}
The following inequality holds for $i\geq 0$:
$$\left|\frac{f^{(n)}(x)}{f(x)}\right|\leq\left[\Sigma_f(x)\right]^n.$$
The proof is a straight-forward computation.  See the proof of \citep[Lemma 6.2]{Burr-Krahmer-Yap:integral:09} or \citep[Section 5.2]{sagraloff-yap:ceval:09} for details.
\end{lemma}

We use this lemma to show that a simple upper bound on the width of an interval will ensure that the conditions in the \EVAL\ algorithm hold. For example, in condition $\text{C}_0$, divide both sides of the inequality by $|f(m(J))|$ and apply Lemma \ref{lem:Sigma} to derive the following inequality:
$$\sum_{i=1}^d\frac{|f^{(i)}(m(J))|}{i!|f(m(J))|}\left(\frac{w(J)}{2}\right)^i\leq\sum_{i=1}^d\frac{1}{i!}\left(\frac{\Sigma_f(m(J))w(J)}{2}\right)^i.$$
If $w(J)\leq\frac{1}{\Sigma_f(m(J))}$, then the sum on the RHS is bounded above by a geometric series with $r=1/2$, and, therefore, the sum is bounded by 1.  This implies that condition $\text{C}_0$ holds.  Therefore, the condition $w(J)\leq\frac{1}{\Sigma_f(m(J))}$ is sufficient to ensure that $J$ is \EVAL\ terminal.  Similarly, if $w(J)\leq\frac{1}{\Sigma_{f'}(m(J))}$, then $J$ is \EVAL\ terminal by condition $\text{C}_1$.

\section{Stopping functions}\label{Sec:Integral}
In this section, we show how stopping functions can be used to compute the size of the subdivision tree of the \EVAL\ algorithm.  The construction in Section \ref{sec:BasicProperties} was originally presented in \citep{Burr-Krahmer-Yap:integral:09}, but we include it here for completeness and because the construction in Section \ref{sec:stopeval} requires a detailed understanding of the method.

\subsection{Basic properties}\label{sec:BasicProperties}
The use of stopping functions promises to be an important tool for bounding the complexity of subdivision algorithms.  Most of the numerical algorithms appearing in the introduction may benefit from this type of analysis; more algorithms of this type are mentioned in the Conclusion, Section \ref{Conc}.  We begin by formulating an abstract algorithm called the \GEN\ algorithm, which is intended to be the prototype of these types of algorithms in one dimension.  The notion of stopping functions and the \GEN\ algorithm both easily generalize to higher dimensions.

Fix a predicate $B$ (i.e., a Boolean function) on intervals with the following property: if $K\subseteq J$ and $B(J)$ is true, then $B(K)$ is also true.  The \GEN\ algorithm is the following algorithm: given an interval $I$, the algorithm maintains a partition $P$ of $I$.  Initially, let the partition be the trivial partition $P=\{I\}$ and let $P_{\GEN}(I)$ be the final partition.\\[.2cm]
\fbox{\parbox{4.7in}{
\begin{center}
Algorithm 3.1: The \GEN\ algorithm
\end{center}
\it
Repeatedly subdivide each $J\in P$ until the following condition holds:\\[.15cm]
\hspace*{40pt} $B(J)$ is true
}}\\

A {\em stopping function} for the \GEN\ algorithm with predicate $B$ is a real-valued function $F$ with the following property: if, for a given interval $J$, there exists a point $p\in J$ such that $w(J)\leq F(p)$, then $B(J)$ is true.  The following theorem, which also appears as \citep[Theorem 3.5]{Burr-Krahmer-Yap:integral:09}, bounds the number of subdivisions performed by the \GEN\ algorithm.

\begin{theorem}\cite[Theorem 3.5]{Burr-Krahmer-Yap:integral:09}\label{thm:integral}
Let $F$ be a stopping function for the \GEN\ algorithm, then
$$\#P_{\GEN}(I) \leq\max\left\{1,\int_I\frac{2dx}{F(x)}\right\}.$$
If the \GEN\ algorithm does not terminate, then the integral is infinite.
\begin{proof}
If $\#P_{\GEN}=1$, then the bound is immediate.  If $\#P_{\GEN}>1$, then an examination of the \GEN\ algorithm shows that for $J\in P_{\GEN}$ there is a lower bound on $w(J)$ since the \GEN\ did not terminate at the parent of $J$:
$$\forall c\in J, w(J)\geq\frac{1}{2} F(c).$$
In addition, $\int_I{\frac{2dx}{F(x)}}=\sum_{J\in P_{\GEN}}\int_J{\frac{2dx}{F(x)}}$, and it, therefore, suffices to show that for every $J\in P_{\GEN}$, $\int_J{\frac{2dx}{F(x)}}\ge1$.  Let $d\in J$ be such that $F(d)$ is maximal in $J$.  Then
\begin{equation*}
\int_J{\frac{2dx}{F(x)}}\ge\int_J{\frac{2dx}{F(d)}}=\frac{2}{F(d)}w(J)\ge\frac{2}{F(d)}\cdot\frac{F(d)}{2}=1.
\end{equation*}
In the case when the \GEN\ algorithm does not terminate, we can look at the partition $P$ at any moment in time.  The above argument shows that $\#P$ is still bounded by the integral $\int_I 2dx/F(x)$.  Since $\#P$ can be chosen to be arbitrarily large, this shows that the integral is unbounded.
\end{proof}
\end{theorem}

\subsection{A stopping function for \EVAL}\label{sec:stopeval}
The next goal is to transform the inequality $w(J)\leq\frac{1}{\Sigma_f(m(J))}$ into a stopping function.  Currently, it is not a stopping function because the function on the RHS is not for an arbitrary point of $J$, but for a specific point, the midpoint.  We begin to turn this into a stopping function via the following lemma:

\begin{lemma}\label{lem:HM}
Let $\overline{z}=(z_1,\cdots,z_d)$ with $z_i>0$ and $y\in\mathbb{R}$ such that $y>0$ and $z_i>y$ for all $i$.  Then
\begin{align}
\label{Eq:HM1}\HM(\overline{z}-y)&\geq \HM(\overline{z})-d\cdot y\\
\label{Eq:HM2}\HM(\overline{z})&\leq d\cdot z_i\qquad\forall i.
\end{align}
\begin{proof}
For Inequality (\ref{Eq:HM1}), we expand each of the harmonic means and get the following equivalent inequality:
$$\frac{d}{\sum\frac{1}{z_i-y}}\geq\frac{d}{\sum\frac{1}{z_i}}-d\cdot y.$$
Noting that all of the denominators are positive, clearing fractions gives that this inequality is equivalent to the following inequality:
$$y\left(\sum\frac{1}{z_i-y}\right)\left(\sum\frac{1}{z_i}\right)\geq \sum\frac{1}{z_i-y}-\sum\frac{1}{z_i}.$$
This inequality is easily justified by combining similar terms on the RHS to obtain a sum with general term
$$\frac{1}{z_i-y}-\frac{1}{z_i}=\frac{y}{z_i(z_i-y)},$$
which is a term that appears on the LHS.  Since the remaining terms on the LHS are positive, this proves the first inequality.

For Inequality (\ref{Eq:HM2}), we expand the harmonic mean to get the following equivalent inequality:
$$\frac{d}{\sum\frac{1}{z_i}}\leq d\cdot z_i.$$
Once again, the denominator is positive, so by clearing fractions we have that this inequality is equivalent to the following inequality:
$$\frac{1}{z_i}\leq\sum\frac{1}{z_i}.$$
Since all of the terms on the RHS are positive and include the term on the LHS, this proves the second inequality.
\end{proof}
\end{lemma}

Let $G_0(x)=\frac{2}{3\Sigma_f(x)}$, then $G_0$ is a stopping function for EVAL: Let $J$ be an interval such that $J$ contains $x$ and let $m$ be the midpoint of $J$; then, $|x-m|\leq \frac{w(J)}{2}$.  Assume now that $w(J)\leq\frac{2}{3\Sigma_f(x)}$.  Then, inequality (\ref{Eq:HM2}) in Lemma \ref{lem:HM} implies that $|x-\alpha|\geq\frac{1}{\Sigma_f(x)}>\frac{w(J)}{2}$ for all $\alpha\in\alpha_f$.  This setup implies the following inequalities:
$$w(J)\leq\frac{1}{\Sigma_f(x)}-\frac{w(J)}{2} \leq\frac{\HM(|x-\alpha_f|-\frac{w(J)}{2})}{d} \leq\frac{\HM(|x-\alpha_f|-|x-m|)}{d}
\leq\frac{1}{\Sigma_f(m)}.$$
The second inequality follows from Lemma \ref{lem:HM} and the fact that the terms of the harmonic mean $\HM(|x-\alpha_f|-|x-m|)$ are all positive (because of the bound on $w(J)$ above).  The remaining inequalities follow from the monotonicity of the harmonic mean.  The last inequality also uses that $|x-\alpha_j|-|x-m|\leq|m-\alpha_j|$ by the triangle inequality.  When combined with the observations from Section \ref{sec:evaltermint}, this implies that $J$ is \EVAL\ terminal.  Similarly, let $G_1(x)=\frac{2}{3\Sigma_{f'}(x)}$ where $\Sigma_{f'}$ is the corresponding function for $f'$, then $G_1$ is also a stopping function for the \EVAL\ algorithm.  Finally, let $G(x)=\max\{G_0(x),G_1(x)\}$, then $G$ is an everywhere positive stopping function for the \EVAL\ algorithm.

\section{Size of the subdivision tree of the \EVAL\ algorithm for the benchmark problem}\label{Sec:Analysis}
In this section, we prove that the size of the subdivision tree of the \EVAL\ algorithm is $O(d(L+\ln d))$ where $L$ is the number of bits needed to write the coefficients of $f$.  In this case, the absolute value of all the roots is bounded by $2^L$ \citep{yap:algebra:bk} (this bound comes from the original $f$, not from the square free substitution).  Hence, we can assume wlog that $b=-a=2^L$.  By Theorem \ref{thm:integral}, the complexity of the \EVAL\ algorithm is bounded by $\int_I\frac{2}{G(x)}dx$.  The crossover points of $G$ are difficult to determine, however, so we replace this integral by a slightly larger one which is easier to evaluate: For any $x\in I$, let $R_x$ be the set of roots in $\alpha_{f\! f'}$ which are closest to $x$.  Similarly, for $\alpha\in \alpha_{f\! f'}$, let $I_\alpha$ be the set of $x\in I$ such that no other root in $\alpha_{f\! f'}$ is closer to $x$ than $\alpha$.  Note that $x\in I_\alpha$ iff $\alpha\in R_x$ and that two of the $I_\alpha$'s are either disjoint (except for endpoints) or coincide (in the case of complex conjugates).  Therefore, these $I_\alpha$'s determine a partition of $I$.  Also, let $S$ be the set of endpoints of the $I_\alpha$'s; then, for all points $x\in I\setminus S$, one has $R_x\subseteq\alpha_f$ or $R_x\subseteq\alpha_{f'}$ because $f$ and $f'$ do not share roots.  We define another function $F(x)$:
$$F(x)=\begin{cases}G_1(x)&x\not\in S\text{ and }R_x\subseteq\alpha_f\\
G_0(x)&x\not\in S\text{ and }R_x\subseteq\alpha_{f'}\\
G(x)&x\in S\end{cases}.$$
Note that although $S$ might not correspond to the crossover points of $G$, pointwise, $F(x)\leq G(x)$ since $G$ is a maximum of the terms which can occur in $F$.  This implies the following inequalities:
\begin{equation}\label{eq:evaluated}
\int_I\frac{2}{G(x)}dx\leq\int_I\frac{2}{F(x)}dx\leq \int_I\sum_{\alpha\in \alpha_{f\! f'}\setminus R_x}\frac{3dx}{|x-\alpha|}=\sum_{\alpha\in \alpha_{f\! f'}}\int_{I\setminus I_{\alpha}}\frac{3dx}{|x-\alpha|}.
\end{equation}
For the second inequality let $x\not\in S$, then $x$ is either closest to a root of $f$ or a root of $f'$. If $x$ is closest to a root of $f$, then $R_x\subseteq\alpha_f$ and $\frac{2}{F(x)}=3\Sigma_{f'}(x)$.  In this case, the sum to the right of the inequality includes all of the roots in $\alpha_{f'}$ as well as some roots in $\alpha_f$.  Thus, at least all of the terms of $\Sigma_{f'}(x)$ appear on the RHS of the inequality.  The case where $x$ is closest to a root of $f'$ is similar.  This implies the inequality because the set of points for which this inequality may fail is a measure zero subset of $S$.

\subsection{Evaluating the integrals}
Consider the shape of each of the regions where we integrate: since all of the integrals are of the form $\int_r^s \frac{3dx}{|x-\alpha|}$, we evaluate a general integral of this form where $r$ and $s$ lie on the same side of $\Real(\alpha)$.
\begin{itemize}
\item In the case where $\alpha$ is real:
\begin{align*}
\text{if }&s>r>\alpha&\text{if }&r<s<\alpha\\
\int_r^s&\frac{3dx}{|x-\alpha|}=\int_r^s\frac{3dx}{x-\alpha}&\int_r^s&\frac{3dx}{|x-\alpha|}=\int_r^s\frac{3dx}{\alpha-x}\\
&\quad=3\ln(|s-\alpha|)-3\ln(|r-\alpha|)&&\quad=3\ln(|r-\alpha|)-3\ln(|s-\alpha|)
\end{align*}
These logarithms will be bounded in the next section.
\item In the case where $\alpha$ is not real:
\begin{align*}
\int_r^s\frac{3}{|x-\alpha|}dx&=\int_r^s\frac{3}{\sqrt{(x-\Real(\alpha))^2+\Imag(\alpha)^2}}dx\\
&=\int_{(r-\Real(\alpha))/|\Imag(\alpha)|}^{(s-\Real(\alpha))/|\Imag(\alpha)|}\frac{3}{\sqrt{y^2+1}}dy\\
&=3\arcsinh\left(\frac{s-\Real(\alpha)}{|\Imag(\alpha)|}\right)-3\arcsinh\left(\frac{r-\Real(\alpha)}{|\Imag(\alpha)|}\right)
\end{align*}
This is now bounded via the relationship between $\Real(\alpha)$ and $r,s$.  If $s>r>\Real(\alpha)$, then:
\begin{align*}
3\arcsinh&\left(\frac{s-\Real(\alpha)}{|\Imag(\alpha)|}\right)-3\arcsinh\left(\frac{r-\Real(\alpha)}{|\Imag(\alpha)|}\right)\\
&=3\ln\left(\frac{s-\Real(\alpha)}{|\Imag(\alpha)|}+\sqrt{\left(\frac{s-\Real(\alpha)}{|\Imag(\alpha)|}\right)^2+1}\right)\\
&\hspace{50pt}-3\ln\left(\frac{r-\Real(\alpha)}{|\Imag(\alpha)|}+\sqrt{\left(\frac{r-\Real(\alpha)}{|\Imag(\alpha)|}\right)^2+1}\right)\\
&=3\ln(s-\Real(\alpha)+\sqrt{(s-\Real(\alpha))^2+\Imag(\alpha)^2})\\
&\hspace{50pt}-3\ln(r-\Real(\alpha)+\sqrt{(r-\Real(\alpha))^2+\Imag(\alpha)^2})\\
&\leq3\ln(2|s-\alpha|)-3\ln(|r-\alpha|).
\end{align*}
If $r<s<\Real(\alpha)$, then the computation is similar, and the integral is bounded above by $3\ln(2|r-\alpha|)-3\ln(|s-\alpha|)$.  These logarithms will also be bounded in the next section.
\end{itemize}

\subsection{Finishing the bound on the \EVAL\ algorithm}\label{subsec:EVALbounds}
In this section, we use the computation from the previous section to prove the main result of this paper.  To do this, we consider the roots $\alpha\in\alpha_{f\! f'}$ with two different cases depending on if $\alpha$ is real or not.
\begin{itemize}
\item If $\alpha$ is real; then $\alpha\in I_\alpha$ and let $I_\alpha=[c,d]$.  Then, the term corresponding to $\alpha$ in the RHS of Inequality (\ref{eq:evaluated}) consists of $\int_{I\setminus I_{\alpha}}\frac{3dx}{|x-\alpha|}=\int_{-2^L}^c\frac{3dx}{|x-\alpha|}+\int_d^{2^L}\frac{3dx}{|x-\alpha|}$.  Note that integrals may be zero which happens when $c=-2^L$ or $d=2^L$.  Then, using the bounds derived in the preceding section on these integrals, it follows that they are bounded by:
    $$3\ln(|-2^L-\alpha|)-3\ln(|c-\alpha|)+3\ln(|2^L-\alpha|)-3\ln(|d-\alpha|).$$
    The positive terms are bounded by $O(L)$ (the leading term is $6\ln (2)L$) and for the negative terms, note that $c$ and $d$ are points which are equidistant from $\alpha$ and another root of $f\!f'$, e.g., $c$ is equidistant from $\alpha$ and $\beta\in\alpha_{f\!f'}$ where $I_\beta$ is the interval immediately to the left of $I_\alpha$.  Then, $\ln(|c-\alpha|)$ is bounded below by the logarithm of half the distance from $\alpha$ to $\beta$.
\item If $\alpha$ is not real, then the term corresponding $\alpha$ in the RHS of Inequality (\ref{eq:evaluated}) consists of $\int_{I\setminus I_{\alpha}}\frac{3dx}{|x-\alpha|}$ which is bounded above by $\int_I\frac{3dx}{|x-\alpha|}$.  By splitting this integral at $\Real(\alpha)$, the integral is equal to $\int_{-2^L}^{\Real(\alpha)}\frac{3dx}{|x-\alpha|}+\int_{\Real(\alpha)}^{2^L}\frac{3dx}{|x-\alpha|}$. Using the bounds derived in the preceding section on these integrals, it follows that these integrals are bounded by:
    $$3\ln(2|-2^L-\alpha|)-3\ln(|\Real(\alpha)-\alpha|)+3\ln(2|2^L-\alpha|)-3\ln(|\Real(\alpha)-\alpha|).$$
    The positive terms are bounded by $O(L)$ (the leading term is $6\ln(2)L$) and for the negative terms, note that $|\Real(\alpha)-\alpha|=|\Imag(\alpha)|$, which is the logarithm of half the distance between $\alpha$ and $\overline{\alpha}$.
\end{itemize}
Combining all of the $O(L)$'s which appear in the integrals results in a bound of $O(dL)$ (the leading term is $6(\ln 2(2d-1))L$).  The sum of the logarithmic distances between roots are bounded simultaneously via the standard Mahler-Davenport lower bound on distances between roots, see \citep{davenport:85,du-sharma-yap:sturm:07,eigenwillig-sharma-yap:descartes:06}.  To do this, we construct a directed graph whose nodes are the roots in $\alpha_{f\!f'}$ and whose edges represent the logarithms which must be calculated.  In this graph, the edges satisfy the conditions of the Mahler-Davenport bound and are chosen so that the in-degree of any node is at most 2.  For each pair of complex roots, $(\alpha,\overline{\alpha})$, connect them with two directed edges, one from $\alpha$ to $\overline{\alpha}$, the other in the opposite direction.  On the other hand, if $\alpha$ is real, then let $\beta$ be a root where $I_\beta$ lies immediately to the right of $I_\alpha$ and $\gamma$ be a root where $I_\gamma$ lies immediately to the left of $I_\alpha$ (provided $I_\alpha$ is not the rightmost or leftmost interval in the partition, respectively).  Those of $\beta$ or $\gamma$ which are real are connected to $\alpha$ so that the arrow points in the direction of decreasing absolute value.  If $\beta$ is not real, then connect $\alpha$ to either $\beta$ or $\overline{\beta}$, whichever has {\em positive} imaginary component.  On the other hand, if $\gamma$ is not real, then connect $\alpha$ to either $\gamma$ or $\overline{\gamma}$, whichever has {\em negative} imaginary component.  Again, these edges are directed so that the arrow points in the direction of decreasing absolute value.  By inspection, we find that the maximum in-degree of this directed graph is $2$.  The Mahler-Davenport bound can then be applied twice to find the result.  The bound implies that the sum of the negative logarithmic distances between the roots appearing in this construction is bounded above by:
$$12\cdot\ln\left(\frac{1}{\sqrt{|\disc(f\!f')|}}M(f\!f')^{2d-2}\left(\frac{2d-1}{\sqrt{3}}\right)^{2d-1}(2d-1)^d\right).$$
The discriminant will be an integer and therefore the discriminant term is bounded above by 1.  The Mahler measure of $f\!f'$ is bounded in terms of the $2$-norms of the coefficients of the original $f$ and $f'$: $M(f\!f')=M(f)M(f')\leq \|f\|_2\|f'\|_2\leq (2^L\sqrt{d+1})(d2^L\sqrt{d})$.  Therefore, this portion is bounded by $O(dL+d\ln d)$ (the leading term is bounded $24\ln(2)dL+42d\ln d)$).  Thus, the complexity of the \EVAL\ algorithm is $O(d(L+\ln d))$ (the leading term is bounded $36\ln(2)dL+42d\ln d\leq 25dL+42d\ln d$).

If $f$ or $f'$ was replaced by a square free version, we used the original $f$ and $f'$ because the square free versions of $f$ and $f'$ divide the original functions, and, therefore, the Mahler measure of the product of the square free versions is bounded above by $M(ff')$.  In fact, the $2$-norms of the coefficients of the original functions are often smaller than the $2$-norms of the square free versions.
\section{Conclusion}\label{Conc}
In this paper, we provided a complexity analysis of the \EVAL\ algorithm and showed it to be optimal under the weak assumption that $L\geq \ln d$.  To accomplish this, we used the novel technique of continuous amortization through stopping functions.  The simplicity of this argument exhibits the utility of this technique: the proof of the next closest complexity bound for an \EVALTYPE\ algorithm in \citep{sagraloff-yap:ceval:09} is significantly more complex.

The \EVAL\ algorithm is very easy to implement \citep{moore:bk,mitchell:robust-ray:90,plantinga-vegter:isotopic:04,plantinga:thesis:06,kamath:subdivision:10} and it now joins the Sturm and Descartes methods by having a subdivision tree which grows at the rate $O(d(L+\ln d))$.  It, therefore, may become more prevalent in practical situations because it has several desirable properties.  This also answers a question raised in \citep{Henrici:search:70} concerning the good behavior of this technique.

In addition, the continuous amortization technique can be used to bound the number of subdivisions over any interval, and, therefore, may find many more applications for different types of questions about subdivision algorithms.  For example, in many practical applications, the question is to find the roots in a given domain, not just for the benchmark domain; continuous amortization may provide a comparison of different algorithms in these situations.

We close with some continuing research and questions:
\begin{itemize}
\item The algorithm for finding complex roots appearing in \citep{sagraloff-yap:ceval:09} is very similar to the \EVAL\ algorithm. We are currently preparing a simplification of their work using the results from this paper.
\item There are many bisection algorithms where continuous amortization may be useful, see, for example, \citep{Henrici:search:70,Yakoubsohn:bisection:05,sagraloff-yap:ceval:09, plantinga-vegter:isotopic:04,plantinga:thesis:06,snyder:interval:92, galehouse:thesis,burr+3:subdiv2:10,eigenwillig-sharma-yap:descartes:06, du-sharma-yap:sturm:07,linyap2009}.  We plan on extending our techniques to these cases.  In particular, stopping functions which are appropriate for the two dimensional cases treated in \citep{plantinga-vegter:isotopic:04,plantinga:thesis:06,galehouse:thesis} would be very useful because current techniques have not been fruitful in establishing complexity bounds of these algorithms.
\item If $f'$ was not square free, then the test for condition $(C_1)$ in Algorithm~2.1 is based on the square free part of $f'$, not the original function.  The \EVAL\ algorithm, however, will continue to terminate and be correct even when this substitution does not occur, i.e., when the original $f'$ is used.  For this reason, it is likely that the above substitution is extraneous.  For example, in the simplest cases where $f'$ is not square free and the integral in Inequality (\ref{eq:evaluated}) can be calculated by hand, the result is $O(d(L+\ln d))$.
\end{itemize}

\bibliographystyle{plainnat}
\bibliography{Subdivision}

\end{document}